\documentclass[twocolumn,superscriptaddress]{revtex4}
\usepackage{graphicx}
\usepackage{epsfig}
\usepackage{amsmath}
\usepackage{mathrsfs}
\usepackage{amsfonts,amsbsy}
\usepackage{amssymb}
\usepackage{hyperref}
\usepackage{bm}
\usepackage{color}
\usepackage{enumerate}
\usepackage{verbatim}
\usepackage{subfloat}
\usepackage{subfigure}
\usepackage{subcaption} 
\usepackage{caption}
\def\be{\begin{equation}}
\def\ee{\end{equation}}
\def\bea{\begin{eqnarray}}
\def\eea{\end{eqnarray}}

\begin{document}

\title{Eigen-microstate Signatures of Criticality in Relativistic Heavy-Ion Collisions}

\author{Ranran Guo}\
\affiliation{Key Laboratory of Quark and Lepton Physics (MOE) and Institute of Particle Physics, Central China Normal University, Wuhan 430079, China}
\author{Jin Wu}
\affiliation{College of Physics and Electronic Information Engineering, Guilin University of Technology, Guilin 541004, Guangxi, China}
\author{Mingmei Xu}
\email{xumm@ccnu.edu.cn}
\affiliation{Key Laboratory of Quark and Lepton Physics (MOE) and Institute of Particle Physics, Central China Normal University, Wuhan 430079, China}
\author{Xiaosong Chen}
\affiliation{Institute for Advanced Study in Physics, Zhejiang University, Hangzhou 310058, China}
\affiliation{School of Systems Science and Institute of Nonequilibrium Systems, Beijing Normal University, Beijing 100875, China}
\author{Zhiming Li}
\affiliation{Key Laboratory of Quark and Lepton Physics (MOE) and Institute of Particle Physics, Central China Normal University, Wuhan 430079, China}
\author{Zhengning Yin}
\affiliation{Key Laboratory of Quark and Lepton Physics (MOE) and Institute of Particle Physics, Central China Normal University, Wuhan 430079, China}
\author{Yuanfang Wu}
\email{wuyf@ccnu.edu.cn}
\affiliation{Key Laboratory of Quark and Lepton Physics (MOE) and Institute of Particle Physics, Central China Normal University, Wuhan 430079, China}

\begin{abstract} 

We develop the eigen-microstate framework as a new approach to identify criticality in relativistic heavy-ion collisions. We construct the original microstate, defined as the final-state particle fluctuations of a single event. By examining ensembles of such original microstates with and without critical signals, we demonstrate that the corresponding eigen-microstate can extract and reveal the dominant critical mode, with the largest eigenvalue serving as a robust order parameter. This framework avoids equilibrium assumptions and is insensitive to non-critical background, and the approach is directly applicable to RHIC Beam Energy Scan data, offering a powerful new tool in the search for the QCD critical point.

\end{abstract}
\maketitle

%%%%%%%%%%%%%%%%%%%%%%%%%%%%%%%%%%%%%%%%%%%%%%%%%%%%%%%%%%%%%%%%%%%%%%%%%%%%%%
% \section{Introduction}\label{Introduction}

Quantum Chromodynamics (QCD) predicts a deconfinement phase transition (PT) from hadronic matter to quark-gluon plasma~\cite{QGP01}. Lattice QCD calculations indicate that at low baryon chemical potential the transition is a smooth crossover~\cite{crossover}, while QCD-inspired models predict a first-order PT at higher baryon density~\cite{1st-PT01,1st-PT02}, terminating at a second-order critical point (CP)~\cite{Stephanov1,Stephanov2}. Locating the QCD CP and mapping the phase boundary are therefore central goals of current and future relativistic heavy-ion programs, including RHIC BES, CBM, NICA, and HIRFL-CSR~\cite{review}.

The natural order parameter associated with QCD deconfinement is the Polyakov loop, but its direct connection to experimentally accessible observables remains unclear. Consequently, theoretical studies have proposed a variety of critical-sensitive probes inspired by statistical physics and QCD, such as higher moments of conserved charges~\cite{Stephanov2,moment02} and factorial moments of multiplicity distributions~\cite{fac-moment,WuYF-PRL}. Over the past decades, extensive efforts have been devoted to measuring higher-order moments of net charge~\cite{net-charge-STAR}, net proton number~\cite{m-STAR-Luo}, and net kaon number~\cite{net-kaon}. However, these observables require very high statistics~\cite{STAR-note,ChenLZ} and careful subtraction of conventional and nonequilibrium effects~\cite{non-eq02,non-eq03,non-eq04,non-eq05,LiXB-PRE}---including trivial statistical fluctuations, resonance decays, and energy-momentum conservation---so the anticipated non-monotonic energy dependence has not yet been conclusively observed~\cite{m-STAR-Luo}. Similarly, factorial-moment analyses, even after non-critical backgrounds are suppressed using mixed-event methods, have revealed only weak intermittency signals at RHIC BES I~\cite{STAR-WuJ,WuJ-2021,WuJ-PRC,WangR,LiZM} and SPS~\cite{NA49}, and a clear power-law scaling has not been established.

These challenges underscore a central difficulty: in heavy-ion collisions, the degree of equilibration is intrinsically uncertain, and the relevant order parameter may not be directly accessible through final-state observables. The recently developed Eigen-Microstate Approach (EMA) is designed precisely for such situations~\cite{ChenXS-01,ChenXS-02,ChenXS-03,ChenXS-04,ChenXS-05}. Its key conceptual step is the generalization of the microstate description from Gibbs-ensemble theory to arbitrary event ensembles. In this formulation, each event is treated as a microstate of the system without assuming thermal equilibrium, stationarity, or any specific microscopic dynamics. The ensemble is thus defined entirely by the event sample itself, making the framework applicable to strongly dynamical, rapidly evolving systems such as relativistic heavy-ion collisions.

Within this generalized ensemble, EMA, as well as previously known principal component analysis (PCA)~\cite{WangLei-PRB}, extracts the dominant collective modes encoded in the event sample and identifies their emergence or ``condensation" closely analogous to Bose–Einstein condensation in which a single mode becomes macroscopically occupied~\cite{Bose-Einstein}. Importantly, this condensation behavior arises from the structure of event-by-event fluctuations, or the critical long-range correlations, not from conventional short-range correlations. EMA has demonstrated its ability to reveal critical patterns and robust order parameters in a wide range of disparate systems—including the Ising~\cite{WangLei-PRB,ChenXS-02} and Vicsek models~\cite{ChenXS-03}, the Kármán vortex street~\cite{ChenXS-04}, and phase transitions in atmospheric and social dynamics---highlighting its versatility and independence from equilibrium assumptions.

In this Letter, we apply the EMA to relativistic heavy-ion collisions. We define the microstate of each event through fluctuations of final-state charged particles in transverse-momentum space. Using Ultra-relativistic Quantum Molecular Dynamics (UrQMD)~\cite{urqmd01,urqmd02} and Critical Monte Carlo (CMC)~\cite{CMC01,CMC02} simulations, we generate event samples with and without critical signals and demonstrate that the EMA can identify not only the presence of critical patterns but also their strength.

%%%%%%%%%%%%%%%%%%%%%%%%%%%%%%%%%%%%%%%%%%%%%%%%%%%%%%%%%%%%

In relativistic heavy-ion collisions, the experimentally accessible observables are the phase-space distributions of final-state charged particles. Even under identical macroscopic conditions---collision energy, nuclear species, and impact parameter---the final-state particle distributions fluctuate significantly from event to event. These variations arise from differences in the dynamical evolution of each collision. When a quark-gluon plasma forms, distinct evolution paths correspond to different times of chemical and kinetic freeze-out. Consequently, each event can be considered a specific original microstate (OM) that encapsulates a unique temporal sequence of the evolution. 

\begin{figure*}[htbp]
     \centering
     \includegraphics[scale=0.8]{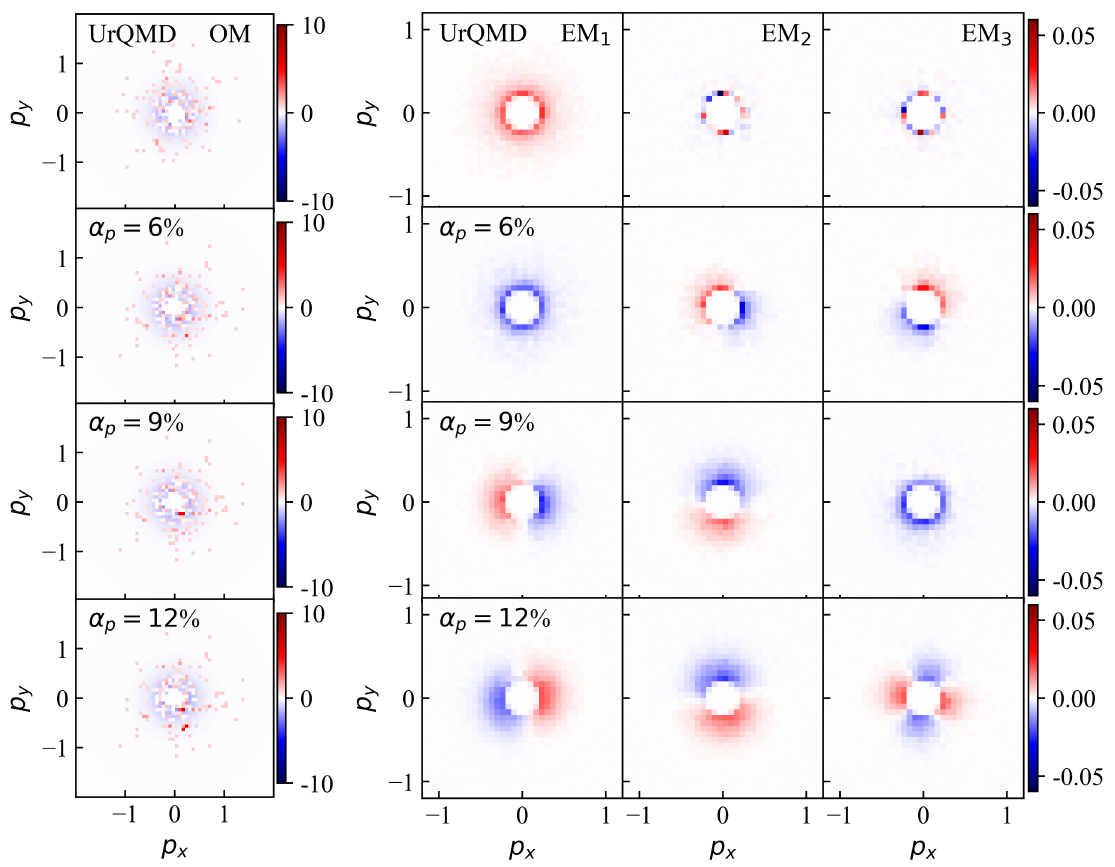}
      \caption{An original microstate (first column) and the top three eigen microstates (columns 2--4) for the UrQMD sample (first row) and corresponding hybrid UrQMD$+$CMC samples with critical-signal fractions $\alpha_{\rm p}=6$\%, 9\%, and 12\% (rows 2--4) at $L=60$.}
     \label{Fig1}
\end{figure*}

To characterize the original microstates, we analyze fluctuations of charged-particle multiplicity in phase-space cells, which are highly sensitive to fractal criticality. Following experimental practice with factorial moments~\cite{NA49,NA61,STAR-WuJ}, the transverse-momentum plane is partitioned into $N=L\times L$ equal lattices by binning $p_x$ and $p_y$ into $L$ segments each. To accommodate the exponential falloff of transverse momentum spectra observed experimentally, both $p_x$ and $p_y$ are restricted to $[-2.0,2.0]$ GeV$/c$. The fluctuation in the 
$i$-th cell of the $I$-th event is defined relative to the event-averaged multiplicity,

\begin{equation}
\Delta N_{{\rm ch},i}^I=N_{{\rm ch},i}^I-\langle N_{{\rm ch},i}\rangle,
\end{equation}
where $N_{{\rm ch},i}^I$ is the number of charged particles and 
$\langle N_{{\rm ch},i}\rangle=\frac{1}{M}\sum_{I=1}^M N_{{\rm ch},i}^I$
is the event-averaged number of charged particle in the cell. The OM vector is constructed from these fluctuations, i.e.,

\begin{equation}
\begin{array}{l}
{\pmb A}^{I}=\frac{1}{\mathscr{N}} \begin{bmatrix}
\Delta N_{{\rm ch},1}^I \\
\Delta N_{{\rm ch},2}^I \\
\vdots \\
\Delta N_{{\rm ch},N}^I
\end{bmatrix}
\end{array},
\end{equation}
where $\mathscr{N}= \sqrt{\sum_{I=1}^M\sum_{i=1}^N(\Delta N^I_{{\rm ch},i})^2}$ is the normalization constant. 

From the ensemble of OMs, we build the temporal-spatial correlation matrix, 

\begin{equation}
    C_{IJ}=[\pmb{A}^I]^T\cdot\pmb{A}^J,\quad I,J=1,2,\dots,M,
\end{equation}
which has dimension $M\times M$, where $M$ is the number of events. Its eigenvalue equation,  

\begin{equation}
    C\pmb{b}_{I}=\lambda_I \pmb{b}_{I},\quad I=1,2,\dots,M,
\end{equation}
defines invariant principal modes under linear transformations, yielding $M$ ordered eigenvalues $\lambda_1 \ge \lambda_2 \ge \dots \ge \lambda_M$ with corresponding eigenvectors,

\begin{equation}
\begin{array}{l}
{\pmb b}_{I}= \begin{bmatrix}
 b_{1I} \\
 b_{2I} \\
\vdots \\
 b_{M,I}
\end{bmatrix}
\end{array}.
\end{equation}
The eigenvectors define the eigen microstates (EMs),

\begin{equation}
    {\pmb E}^{I}=\sum_{l=1}^{M}b_{lI}\pmb{A}^l,
\end{equation}
whose square of amplitude equal to the associated eigenvalues $\lvert{\pmb E}^{I}\rvert^2=[\pmb{E}^I]^T\cdot\pmb{E}^I=\lambda_I$.
The eigenvalues themselves serve as statistical weights $w_I=\lambda_I$, normalized to unity $\sum_{I=1}^{M}w_I=1$, and quantify the contribution of each EM to the full ensemble. If the largest weight $w_1$ becomes finite in the limit $M \to\infty$, this indicates a condensation of EMs in the ensemble. This condensation is similar to the Bose-Einstein condensation~\cite{Bose-Einstein}. The weight $w_1$ therefore plays the role of an order parameter~\cite{WangLei-PRB,ChenXS-01,ChenXS-02,ChenXS-03,ChenXS-04,ChenXS-05}.

To test the sensitivity of the EMA to critical signals, we construct event samples with and without critical fluctuations using the UrQMD and CMC models. The UrQMD model incorporates all essential noncritical dynamics of relativistic heavy-ion collisions, including collective flow effects~\cite{urqmd01,urqmd02}. In contrast, CMC generates particle momenta via Lévy random walks, producing momentum-space fluctuations with fractal criticality~\cite{WuJ-2021,WuJ-PRC,WangR}.

For events containing a controlled critical component, we first generate 0--5\% most-central Au$+$Au collisions at $\sqrt{s_{NN}}=19.6$ GeV with UrQMD (v3.4). A fraction $\alpha_{\rm p}$ of particles in each event is then replaced by particles from a CMC event, defining $\alpha_{\rm p}$ as the fraction of the critical signal (with other schemes to add the signal also possible). To preserve the transverse-momentum spectrum of the original UrQMD sample, replacements are accepted only if $|p_t^{\rm CMC}-p_t^{\rm UrQMD}|<0.2$ GeV$/c$. STAR kinematic cuts are applied: $|\eta|<0.5$ and $p_t$ window $0.2<p_t<1.6$ GeV$/c$ for $K^{\pm}$ and $\pi^{\pm}$~\cite{STAR-WuJ}, and $0.4<p_t<2.0$ GeV$/c$ for $p$ and $\bar{p}$~\cite{m-STAR-Luo}. 

Figure~\ref{Fig1} presents one OM (first column) and the first three EMs (second to fourth columns) for the UrQMD sample (top row) and three hybrid samples with critical-signal fractions $\alpha_{\rm p}=6$\%, 9\% and 12\% (rows 2--4). Each sample contains $M=20,000$
OMs. In each row, the first column shows a randomly selected OM; red and blue denote positive and negative values of $\Delta N_{{\rm ch}}$, respectively. The three adjacent columns display the three EMs with the largest weights.

In the first column, for the UrQMD baseline, azimuthal fluctuations in a single OM appear random and approximately rotationally symmetric. Radial fluctuations are concentrated at smaller radii, reflecting the transverse-momentum distribution, and the gap near the origin results from the cut $0.2 < p_{\rm t} < 2.0$ GeV$/c$. Hybrid samples with $\alpha_{\rm p}=6$\%, 9\%, and 12\%
generally resemble UrQMD OMs, but a few bins show intensified red regions as $\alpha_{\rm p}$ increases, signaling enhanced local fluctuations.

Critical signals alter EM patterns. For $\alpha_{\rm p}=6$\%, the first EM retains a single-color ring corresponding to the conventional collective mode, whereas the second and third EMs begin to show two-patch ring structures with broken rotational symmetry---evidence of a nascent critical mode. As $\alpha_{\rm p}$ increases to 9\%, the weights shift: the critical-related collective mode advances while the conventional mode recedes, indicating growing critical dominance.

At $\alpha_{\rm p}=12$\%, the conventional-mode weight further decreases, and critical-related EMs exhibit pronounced structures: the first two EMs display two-patch patterns and the third shows a four-patch pattern. 
Higher EMs (not shown) exhibit six- and eight-patch structures. Large-scale coherence reflects an enhanced correlation length, while multi-patch patterns across scales signal critical pattern, analogous to cluster formation in the Ising model~\cite{WangLei-PRB,ChenXS-01,ChenXS-02}. In comparison, the  UrQMD sample shows a ring-like first EM with uniform positive fluctuations, similar to the distribution of mean charged multiplicity, while its second and third EMs retain random red-blue distributions along the ring, preserving rotational symmetry and lacking critical cluster like pattern.

\begin{figure}[b]
     \centering
     \includegraphics[scale=0.5]{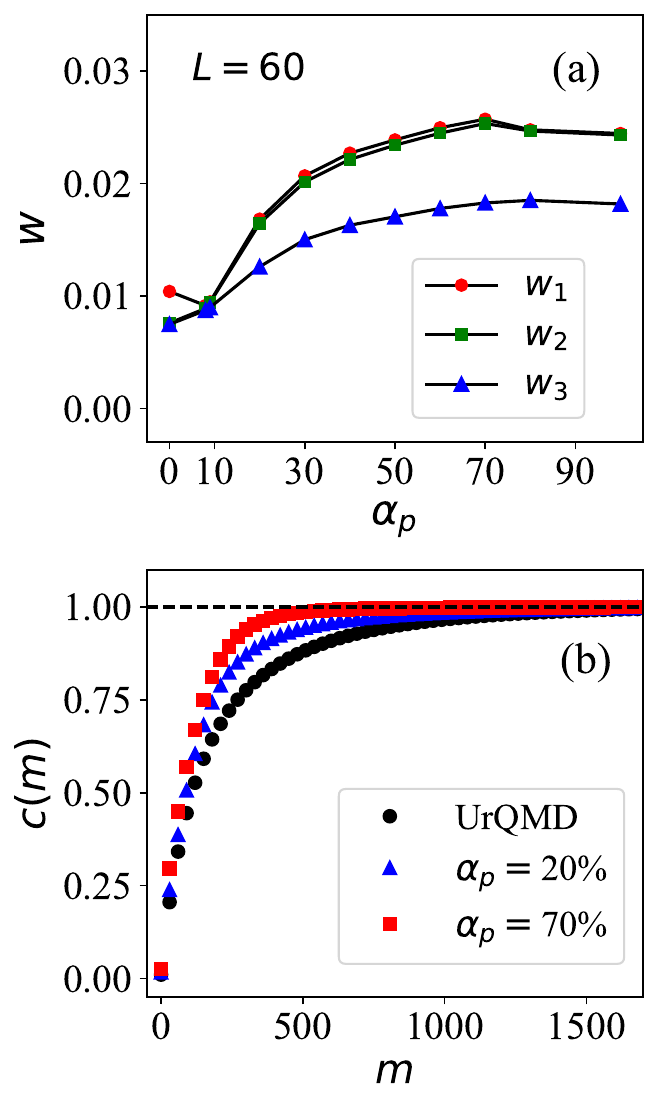}
      \caption{(a) Top three weights $w_{1,2,3}$ as functions of signal fraction $\alpha_{\rm p}$ for $L=60$. (b) Weight cumulants for the original UrQMD sample and hybrid UrQMD$+$CMC samples with $\alpha_{\rm p}=20\%$ and $70\%$.}
      \label{Fig2}
\end{figure}

The emergence of critical patterns with 2-, 4- and 6-patch structures signals a change of order. The bigger the weight, the more dominant the collective mode, the more ordered the system. The weight of collective modes can thus serve as an order parameter, similar to magnetization in the Ising model~\cite{WangLei-PRB,ChenXS-01,ChenXS-02}. A larger critical-signal fraction produces greater weights and a more ordered system. Figure~\ref{Fig2}(a) shows the top three weights, $w_{1,2,3}$, as functions of the signal fraction $\alpha_{\rm p}$. Except for a slight initial drop in $w_1$, all weights increase with $\alpha_{\rm p}$. This initial decrease reflects that, at very small signal fractions, the new component competes with the original non-critical background and is insufficient to form a sizable new phase. As $\alpha_{\rm p}$ increases, critical modes emerge once $\alpha_{\rm p} > 9\%$. Both $w_1$ and $w_2$ peak near $\alpha_{\rm p} \approx 70\%$ and then saturate.

The weight cumulant is defined~\cite{ChenXS-01} as
\begin{equation}
    c(m)=\sum_{I=1}^{m}w_{I}.
\end{equation}
Figure~\ref{Fig2}(b) plots the weight cumulants for the original UrQMD sample and hybrid samples with $\alpha_{\rm p}=20\%$ and $70\%$. In all cases, the cumulant $c(m)$ rises with $m$ and approaches unity. Samples with higher $\alpha_{\rm p}$ saturate earlier, indicating that fewer leading EMs dominate---analogous to the shift from disordered to ordered phases in the Ising model as temperature decreases past criticality~\cite{ChenXS-01,WangLei-PRB}.

\begin{figure}[t]
     \centering
     \includegraphics[scale=0.68]{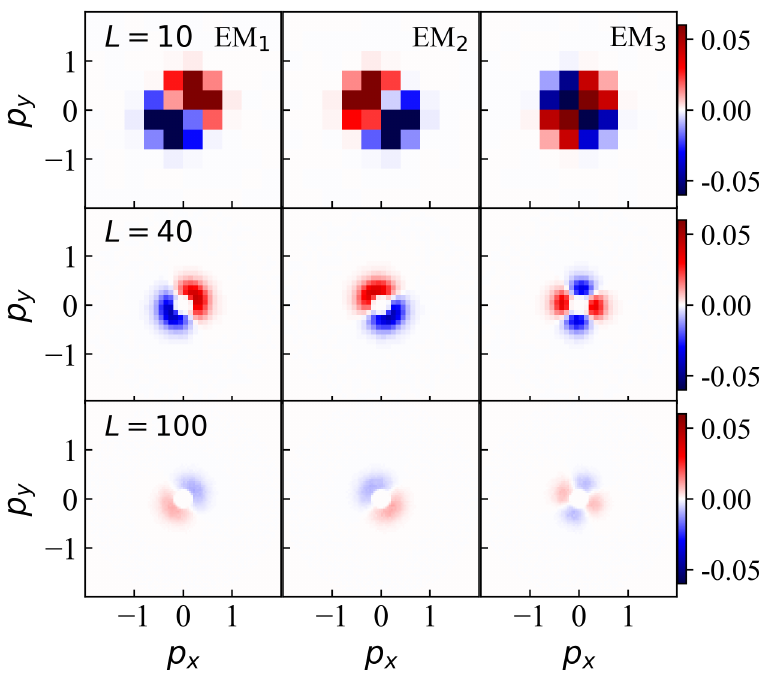}
      \caption{Top three eigen microstates for hybrid UrQMD$+$CMC samples with $\alpha_{\rm p}=70\%$ at $L=10$, 40, and 100. }
     \label{Fig3}
\end{figure}

To confirm the fractal feature of the extracted critical mode, we repeat the analysis for varying division numbers $L$. Figure~\ref{Fig3} displays the three largest EMs in $\alpha_{\rm p}=70\%$ for $L=10,40,100$. For coarse partitioning ($L=10$), grouping adjacent of the same-color reveals two-patch structures in the first two panels and a four-patch pattern in the third---clear signatures of critical patterns even in small division number systems. Increasing $L$ to 40 and 100 leaves the patch structure in the patterns qualitatively unchanged. The lighter colors in larger $L$ reflect reduced fluctuations per cell due to finer binning. This self-similar fractal feature of the patch structure across scales confirms the criticality of the EM pattern.

Meanwhile, to examine the finite-size scaling (FSS) of the corresponding eigenvalue $w_1$, we also plot $w_1$ versus $L$ (double-log scale) around $\alpha_{\rm p}=70\%$, finding a straight-line dependence. The ratios $w_2/w_1$ and $w_3/w_1$ versus $\alpha_{\rm p}$ approach fixed-point–like values for $\alpha_{\rm p}>9\%$. These results further confirm the FSS behavior of $w_1$. So $w_1$ indeed acts as an effective order parameter, similar to magnetization in the Ising model~\cite{ChenXS-01,ChenXS-02,WangLei-PRB}.

%\begin{figure}[t]
%     \centering
%     \includegraphics[scale=0.68]{Fig4.pdf}
%      \caption{$log w_1$ versus $log L$ around $\alpha_{\rm p}=70\%$. }
%     \label{Fig3}
%\end{figure}

The above demonstrations show several advantages of the EMA compared with conventional critical observables in relativistic heavy-ion collisions.
(i) Sensitivity to critical behavior without explicit background subtraction: EMA extracts collective patterns directly from the event ensemble and is less affected by non-critical backgrounds such as statistical fluctuations, resonance decays, and so on.
(ii) No reliance on thermal-equilibrium assumptions: Because the method is formulated in terms of a generalized microstate description, it remains applicable even when equilibration is incomplete, which is expected in the dynamical environment of relativistic heavy-ion collisions.
(iii) Access to an effective order-parameter-like quantity: The largest eigenvalue provides a practical indicator of the strength of the collective mode, offering a potential means to characterize critical behavior when the true order parameter of QCD is experimentally inaccessible.
(iv) Moderate statistical requirements: Stable extraction of eigenvalues and eigen-microstates is achievable with event samples of the order $\sim$ 20,000 microstates, suggesting that the method is computationally efficient compared to higher-order cumulant analyzes. These features indicate that EMA can serve as a useful tool for studying critical patterns in complex, non-equilibrium systems.

% \section{ Summary}\label{Summary}

In summary, we have applied the EMA to relativistic heavy-ion collisions and explored its ability to identify critical behavior in event ensembles. By treating each event as a microstate, the approach isolates dominant colletive modes and their emergence. Analyzes of UrQMD and CMC simulations suggest that the largest eigenvalue acts as an effective indicator of critical pattern formation, while the corresponding eigen-microstate captures characteristic features expected near a phase transition. Since the method does not depend on equilibrium assumptions and appears comparatively robust against non-critical backgrounds, it is in principle directly applicable to RHIC Beam Energy Scan data.

Although further work will be required to assess its full experimental sensitivity, particularly under realistic detector effects and acceptance conditions, the present results indicate that EMA provides a complementary perspective for ongoing searches for critical behavior in RHIC BES-II~\cite{STAR-note} and future heavy-ion programs~\cite{STAR-another}. More generally, the framework offers a systematic way to analyze critical patterns in event-by-event data across a variety of non-equilibrium systems.

%%%%%%%%%%%%%%%%%%%%%%%%%%%%%%%%%%%%%%%%%%%%%%%%%%%%%%%%%%%%% %%%%%%%%%%%%%%%%%%
% \section*{Acknowledgments}

This research was funded by the National Key Research
and Development Program of China, Grant No.
2024YFA1610700, Grant No. 2022YFA1604900, and the
National Natural Science Foundation of China, Grant
No. 12275102 and 12135003.

% The \dots .

\bibliographystyle{apsrev4-2} % 使用 APS 样式
%\onecolumngrid % 切换到单栏模式
\bibliography{ref} 

\end{document}